\documentclass[preprint2]{aastex631}

\shorttitle{X-ray Spectra From Weakly Magnetized Accretion Flows}
\shortauthors{Wallace \& Pe'er}

\graphicspath{{./}{img/}}

\begin{document}

\title{X-ray Spectra From Weakly Magnetized Accretion Flows}

\correspondingauthor{John Wallace}
\email{john.wallace@biu.ac.il, asaf.peer@biu.ac.il}

\author[0000-0001-6180-801X]{John Wallace}
\affiliation{Department of Physics, Bar-Ilan University, Ramat-Gan 52900,
Israel}

\author[0000-0001-8667-0889]{Asaf Pe'er}
\affiliation{Department of Physics, Bar-Ilan University, Ramat-Gan 52900,
Israel}

\begin{abstract}

In this paper, we expand upon previous work that argued for the possibility of a
sub-equipartition magnetic field in the accretion flow of a black hole binary
system. Using X-ray observations of the three well-known sources A0620-00, XTE
J1118+480 and V404 Cyg during the quiescent state, we compare the theoretically
expected spectral shape with the observed data in order to verify that the
parameters of the sub-equipartition model are plausible. In all three cases, we
find that it is possible to reproduce the spectral shape of the X-ray
observations with a sub-equipartition flow. These findings support the idea that
the quiescent state spectrum of X-ray binary systems is produced by a
weakly-magnetized accretion flow. A sub-equipartition flow would pose a
significant challenge to our current understanding of jet-launching, which
relies on the presence of a strong magnetic field to power the jet. 

\end{abstract}


\section{Introduction} \label{sec:intro}

Black hole binaries (BHBs) are luminous sources consisting of a central black
hole, accreting from an orbiting companion star. These sources are generally
observed in one of four main spectral states, named for their X-ray luminosity
and spectral hardness: The high/soft state (HSS), very high state/intermediate
state (VHS/IS), low/hard state (LHS) and quiescent state (QS)
\citep{Esin1997,Zdziarski2004,Remillard2006,Narayan2008}.

BHBs spend most of their lifetime in the extremely low luminosity regime known
as the quiescent state \citep{Belloni2010}. In this state, the low observed
luminosity poses an observational challenge, in spite of the large number of
sources that have been discovered. On the other hand, the influence of a jet on
observations in this state is likely to be greatly reduced \citep{Narayan2008},
making it an ideal laboratory to probe the properties of the accretion flow
itself.

Despite a large number of observed sources \citep[around 75, see][for a recent
survey]{Tetarenko2016}, there is still no clear consensus on the structure of
the accreting system or its radiative properties in each spectral state. The
most common model of such systems consists of two main regions --- an inner, hot
flow which is  optically thin and geometrically thick and an outer,
geometrically thin disk, which is optically thick \citep{Esin1997}. In each of
their spectral states, BHBs are also believed to launch jets from their
innermost regions \citep{Narayan2008}, however the mechanism by which this
occurs is still subject to some debate.

Multiwavelength observations and modeling of X-ray binary spectra play an
important role in developing our understanding of the structure and properties
of the environment surrounding the central black hole \citep[for
example]{Gallo2007,Plotkin2015,Rana2016,Dincer2018}. Understanding the
parameters of the accretion flow in its inner regions is crucial in uncovering
the extreme physics there. In particular, the magnetic field configuration is of
utmost importance in models of jet-launching such as the Blandford-Znajek and
Blandford-Payne mechanisms \citep{Blandford1977, Blandford1982}.

In \citet[hereafter WP21]{Wallace2021}, we showed that the optical/infrared
(OIR) spectra of three quiescent sources could be well explained by the standard
two-component accretion flow, consisting of an outer thin disk, surrounding an
inner, hot accretion flow. In that work, we found that the spectrum at these
frequencies was consistent with a sub-equipartition magnetic field strength in
the inner flow. In addition, radio observations of the three sources also
suggest the presence of a flat-spectrum radio jet \citep{Gallo2007}. These
observations, in combination with the sub-equipartition field found in WP21,
create a tension with our current understanding of jet-launching, which requires
a large magnetic field strength to power and maintain such a jet.

In this paper, we extend our previous work to include X-ray data, in order to
verify that the values used in our previous work are plausible, using the
observed value of the X-ray spectral index. X-ray observations of quiescent BHB
sources generally reveal a power-law slope with spectral index around 2
\citep{Kong2002,Corbel2006,Remillard2006}. This power-law arises naturally from
scattering of soft photons by a population of thermal electrons
\citep{Rybicki1979, Zdziarski1985}. The spectral index of the X-ray spectrum is
tied to the temperature and density of the emitting plasma and therefore we can
use it to confirm that the parameters used to argue for a weak magnetic field in
WP21 are consistent with the observed X-ray spectral slope. This offers further
support for a sub-equipartition field configuration, that not only poses a
challenge for models of jet-launching, but also raises questions about the
dissipation mechanisms necessary to maintain the weakly magnetized state of the
disk.

\section{X-ray Data} \label{sec:data}

We extend the analysis of WP21 using the data described in the following
sections. In each case, we estimate the Compton $y$-parameter and the optical
depth by $y = \left(\nu F_{\nu,\mathrm{IC}}\right) / \left(\nu
F_{\nu,\mathrm{sync}}\right)$ and $\tau = \left(F_{\nu,\mathrm{IC}}\right) /
\left(F_{\nu,\mathrm{sync}}\right)$\citep[see, for example,][]{Pe'er2012}. This
can be used to extract information about the temperature of the plasma, for
example, which allows for a cross-check of our results with the observational
data. The model spectra are constructed according to the procedure described in
WP21.

\subsection{A0620-00}

A0620-00 is a soft X-ray transient source, consisting of a $6.6~M_\odot$ black
hole, orbited by a $0.4~M_\odot$ K-type companion, with an orbital period of
7.75~hr \citep{McClintock1986,Cantrell2010}. It is one of the most extensively
observed BHB systems to date.

Several X-ray observations of A0620-00 in quiescence have been carried out using
Chandra, with observations carried out in 2000, 2005 and 2013, as reported in
\citet{Dincer2018}. The observed spectral index, $\Gamma_\mathrm{obs}$, varies
slightly across different observations, falling in the range $2.07 \pm 0.13 -
2.32 \pm 0.16$. For our purposes, we make use of the 2005 observations,
originally reported by \citet{Gallo2007} (with $\Gamma_\mathrm{obs} = 2.24 \pm
0.16$), which are strictly simultaneous in radio and X-ray, and are nearly
simultaneous with optical data from the Small and Moderate Aperture Research
Telescope (SMARTS). The optical observations were taken one day before the
radio/X-ray data, while infrared data was taken 5 months beforehand (March 2005
vs. August 2005) with Spitzer. This non-simultaneity is important to note when
evaluating the conclusions of our study, given that a variability timescale of
months is typical of BHB systems \citep{Remillard2006}. However, A0620-00 has
not been observed to undergo an outburst between 1996 and 2015
\citep{Tetarenko2016}, combined with the consistent measurements of
$\Gamma_\mathrm{obs}$ over this time frame, we therefore believe it is
reasonable to assume that its radiative properties are relatively stable over
this period of observation.

\subsection{XTE J1118+480}

XTE J1118+480 is another soft X-ray transient source, consisting of a
$7.3~M_\odot$ black hole orbited by a $0.2~M_\odot$ companion, with an orbital
period of 4.1~hr. We use the X-ray observations reported in \citet{Plotkin2015},
which were carried out using Chandra in 2013. The observed spectral index,
$\Gamma_\mathrm{obs}$, is reported to be $2.02 \pm 0.41$.

These observations were carried out nearly-simultaneously with radio, near
infrared (NIR)/optical and ultraviolet (UV) measurements. The radio measurements
were obtained by the Very Large Array (VLA), the NIR/optical by the William
Herschel Telescope (WHT) and the UV by Swift. In addition to the simultaneous
data, there are also several non-simultaneous measurements available for XTE
J1118+480, detailed in \citet{Plotkin2015}.

\subsection{V404 Cygni}

V404 Cyg consists of a $9~M_\odot$ black hole orbited by a $0.7~M_\odot$ K-type
companion, with an orbital period of 155~hr \citep{Bernardini2016}. There are
several X-ray observations of V404 Cyg in its quiescent state, that are all
generally in agreement. Here, we use the observations compiled by
\citet{Hynes2009}, which are primarily based on a simultaneous observing
campaign carried out in 2003, supplemented by archival (non-simultaneous) data.
Radio data were obtained using both the VLA and the Westerbork Synthesis Radio
Telescope (WSRT), the optical measurements are from WHT, the UV from the Hubble
Space Telescope (HST) and the X-ray from Chandra.

\section{Results} \label{sec:results}

\begin{figure} \plottwo{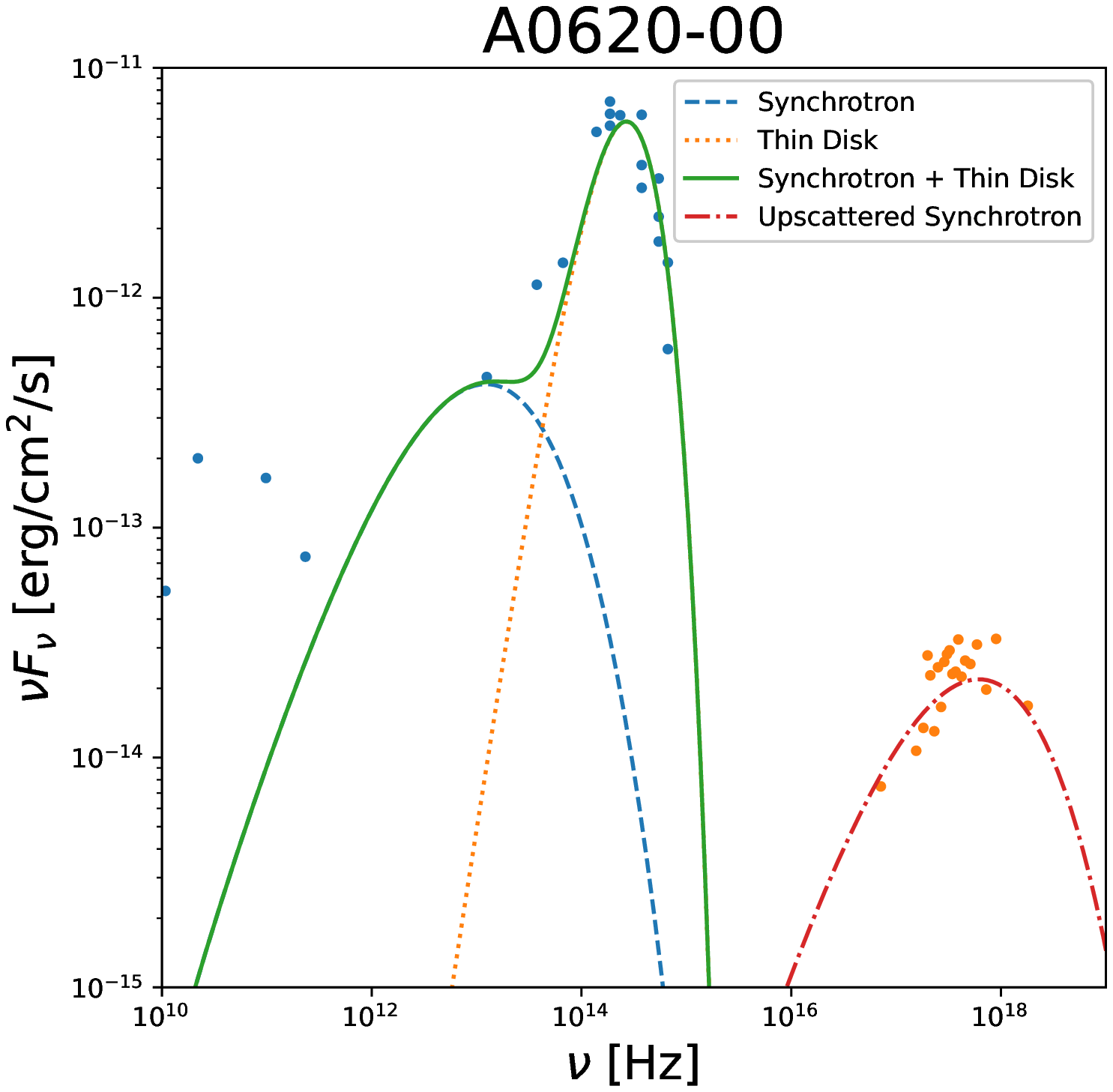}{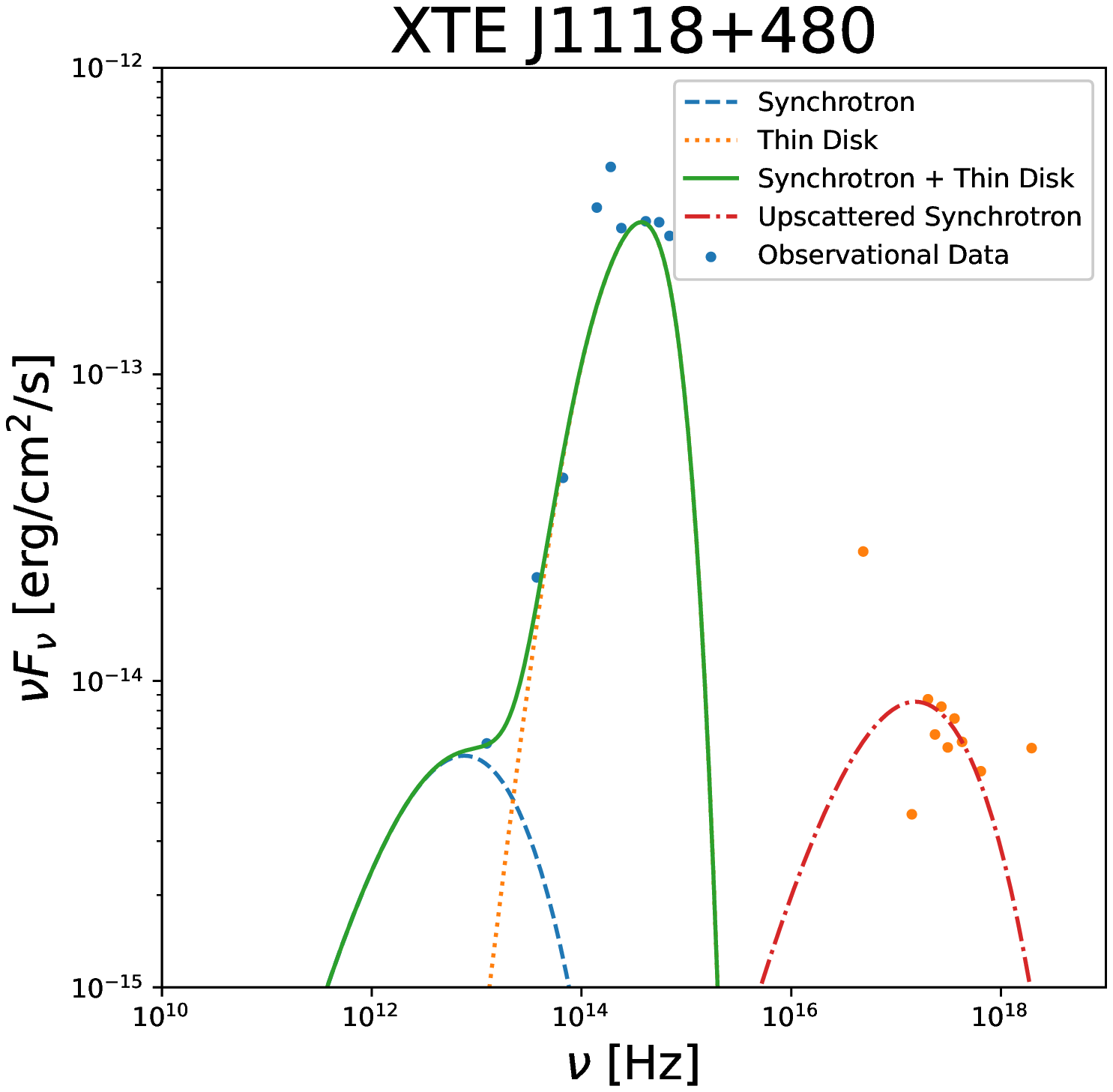} \caption{Spectral energy
   distributions of A0620-00 (left) and XTE J1118+480 (right). These figures are
   similar to those in WP21, but with the addition of X-ray data from Chandra.
   The X-ray emission calculated for the sub-equipartition model shows good
   agreement with the X-ray data in both cases. Typical values of the
   equipartition ratio $\varepsilon_\mathrm{B}$ here are $10^{-8}$, indicating a
   very weakly-magnetized disk.}
   \label{fig:goodsources}
\end{figure}

\begin{figure} \plotone{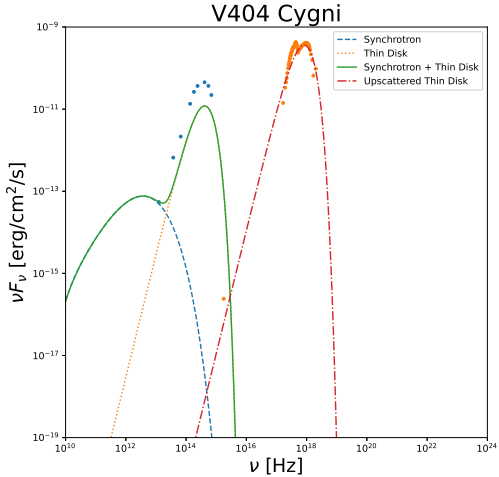} \caption{Spectral energy distribution of V404
    Cyg. This figure is similar to that in WP21, but with the addition of X-ray
    data from Chandra. Here, we have only an upper bound on the emission,
    however under those constraints, the spectral shape from the
    sub-equipartition model again shows good agreement with the X-ray data. In
    this case, it is upscattered disk photons which produce the X-ray emission,
    rather than the synchrotron photons.} \label{fig:v404} \end{figure}

In the cases of A0620-00 and XTE J1118+480 (see fig. \ref{fig:goodsources}), we
find good agreement with the observational data from infrared to X-rays. For
A0620-00, a Compton $y$-parameter of $5\times 10^4$ closely reproduces the
observed X-ray data, corresponding to a plasma optical depth $\tau$ of $\sim
10^{-5}$. In XTE J1118+480, we find that $y = 2\times 10^4$ and $\tau \sim
10^{-5}$ give good agreement with observations. These values are sensible, given
that the synchrotron emission in this model is assumed to come from a very hot,
optically thin flow.

The accretion rate we used for A0620-00 is on the order of $10^{-9} M_\odot
\mathrm{yr}^{-1}$, which corresponds to around 0.5\% of the Eddington value. In
XTE J1118+480, the accretion rate is around $10^{-10} M_\odot \mathrm{yr}^{-1}$,
or $5\times 10^{-4} \dot{M}_\mathrm{Edd}$. The temperature at the inner boundary
of the hot accretion flow in each case is on the order $10^{11}~K$, falling to
around $10^9~K$ at the transition to the thin disk.

In the case of V404 Cyg (fig. \ref{fig:v404}), it was not possible to directly
match the observations with a purely two-component model. This is most likely
due to contamination of the spectrum by the companion star. It was, however,
possible to constrain the accretion rate to below $10^{-8} M_\odot
\mathrm{yr}^{-1}$ (5\% of the Eddington value), which then allows us to carry
out the same analysis on this source. Again, we find that the magnetic field
must be strongly sub-equipartition in order to reproduce the spectral features
seen in observations at OIR frequencies. For the X-ray emission, we find that in
this case, the emission is better explained by upscattering of the thin disk
photons by the hot  inner flow, rather than upscattering of the synchrotron
photons as in the other two sources. This gives good agreement with the X-ray
data for $y = 2\times 10^3$ and $\tau = 1.5\times 10^{-2}$.

Therefore in all cases, we find that the parameters used here are in good
agreement with the available spectral data, offering further support to the
sub-equipartition magnetic field configuration as a potential model for the
emission in BHB sources. In all three cases, the data used here are
simultaneous, or at least nearly-simultaneous across the different frequencies.
This is important due to the time-varying nature of these objects, with typical
variation timescales of hours to days. Thus even observations that occur close
together, but not simultaneously, may exhibit different spectral features
depending on the source and its properties.

\section{Discussion} \label{sec:discussion}

In each case, we found that the X-ray emission in our model is in good agreement
with observed values. This backs up the conclusion of WP21, that the emission
from these sources in their quiescent states (in particular their optical and
infrared emission) can be explained using synchrotron emission from a thermal
population of electrons in a weakly-magnetized inner disk.

Given the extensive observations of A0620-00 and XTE J1118+480 and their
similarity to other quiescent sources, the viability of a sub-equipartition
magnetic field in those sources raises the prospect that such a configuration
may be common across other sources in the quiescent state. The parameters used
in WP21 are typical of weakly accreting binaries in their quiescent states and
so in combination with the available X-ray data, the results of this paper
further support the sub-equipartition model as an explanation for the broadband
spectrum of not just A0620-00 and XTE J1118+480, but potentially quiescent
sources more broadly.

Since BHB sources are readily observed in the X-ray bands, including in
quiescence, it is vital that any model of their emission be consistent with
observations in that range. In each case, the results obtained here show that
the X-ray data supports the sub-equipartition model, thus passing an important
test. An accretion flow with a sub-equipartition magnetic field can therefore
explain the entire observed spectral energy distribution, from optical to X-ray,
assuming that radio emission from a jet exists (for which there is strong
observational evidence).

These findings underscore the fact that the requirement for such a low magnetic
field has important consequences for our understanding of the physical processes
at play in accretion flows. As discussed in WP21, maintaining the magnetic field
at such low values would require a large amount of dissipation throughout the
flow, which may in turn have dynamical or radiative consequences on the
structure of the system. This also raises issues around jet-launching, given
that the magnetic field is central to that process and there is clear evidence
for the presence of at least weak jet emission in each of these sources.

On the other hand, for V404 Cyg, we find that the X-ray emission is better
explained by upscattering of thin disk photons rather than those from the
synchrotron-emitting inner region. the spectral shape of the observed X-ray
emission here is somewhat harder than the standard value of $\sim 2$, but in
line with observations reported by \citet{Kong2002}, \citet{Corbel2006} and
\citet{Hynes2009}. The behaviour of V404 Cyg is unusual compared to the other
two sources considered here, and its higher luminosity has led to some
suggestions that it may be on the boundary between the quiescent state and the
hard state \citep{Gallo2007}.

V404 Cyg is also observed to experience ``hard-only'' outbursts, undergoing one
such outburst in 2015 \citep{Tetarenko2016}, which is after the period of
observations of \citet{Rana2016}, but given the long period of quiescence that
preceded the outburst, from 1989-2015, this may offer an explanation for the
change in spectral parameters. These hard-only outbursts generally do not follow
the standard path along the hardness-intensity diagram, never reaching the soft
state and instead transitioning only between the hard state and quiescence.
 
\citet{Corbel2006} note that V404 Cyg may not soften between the LHS and QS, as
other sources do, but may even harden. This behavior is unusual in BHBs, where a
softening of the X-ray spectrum is generally observed between the LHS and
quiescence. This absence of softening is also observed to occur in a small
subset of sources, suggesting that perhaps there is a sub-population of BHBs
whose accretion structure does not follow the canonical model.
\citet{Corbel2006} suggest that the difference in spectral index may arise from
differences in the mass transfer rate between long- and short-period binaries.

Indeed, V404 Cyg has a much longer orbital period (155~hr) than both XTE
J1118+480 and A0620-00 (4.1~hr and 7.75~hr respectively). The mass transfer rate
(i.e. the accretion rate) is correspondingly higher in V404 Cyg, which is in
line with theoretical expectations for a binary system accreting by Roche lobe
overflow \citep{Frank2002}. This difference arises due to the fact that the mass
transfer may be driven by one of two separate mechanisms: evolutionary expansion
of the donor star --- known as $n$-driven systems --- or angular momentum losses
through gravitational radiation and magnetic braking --- known as $j$-driven.
Due to constraints on the size of the Roche lobe for $n$-driven systems, this
mechanism can only operate in long period systems. \citet{Menou1999} find that
there is a ``bifurcation period'' separating the range of operation of the two
mechanisms, occurring at $P_\mathrm{bif} = 0.5 - 2~\mathrm{days}$. This places
V404 Cyg firmly in the $n$-driven regime and the other two sources in the
$j$-driven regime. Similarly, GRO 1665-40 is another long-period source, with an
orbital period of around 2.5 days \citep{Tetarenko2016}. Its spectral index has
been observed to be around $1.3$ \citep[see]{Corbel2006}, again harder than the
short-period sources considered here.

We therefore conclude that in all three sources considered, the results obtained
here support the findings of WP21, namely a sub-equipartition magnetic field in
the accretion flow. Under this sub-equipartition framework, we are still left
with the challenge of how a jet could be launched by such a weakly-magnetized
disk, as well as how an accretion flow could efficiently dissipate the magnetic
field to maintain such a weak magnetization. As noted in WP21, the flat/inverted
shape of the spectrum in radio observations of BHBs seem to indicate the
presence of a weak jet even at the low luminosities of the quiescent state.
Given that that the Blandford-Znajek mechanism --- which is the usual mechanism
invoked to explain the launching of jets from around black holes --- requires
the presence of a strong magnetic field, it is not clear that it can operate
efficiently in these scenarios.

\section{Conclusions} \label{sec:conclusions} In this paper, we have extended
the analysis of WP21, using X-ray observational data for the three sources
considered in that work to further support the possibility of a
sub-equipartition flow as a model for the observed spectra in quiescent X-ray
binary systems. We find that in each case, the parameters of the accretion flows
used in WP21 are consistent with X-ray observations of those sources,
reproducing the observed X-ray spectral index in each case. This finding further
supports our previous result, namely that the emission can be described by a
weakly-magnetized, two-component accretion flow.

As noted in our previous work however, we require jet emission to adequately
explain the spectrum at radio frequencies, suggesting that jets are still
launched despite this low magnetization. Given the important role played by the
magnetic field in the current understanding of jet-launching, the possible
existence of jets launched by weakly-magnetized accretion flows raises important
questions about the limits of those models, as well as the dissipation processes
necessary within the accretion flow in order to maintain this low magnetic field
strength.

\begin{acknowledgments}
The authors acknowledge support from the European Research Council via the ERC
consolidating grant \#773062 (acronym O.M.J.).
\end{acknowledgments} 

\bibliography{../bibliographies/equipartition-xray}
%\bibliography{equipartition-xray}
\bibliographystyle{aasjournal}

%\listofchanges

\end{document}